\documentclass[twocolumn,showpacs,showkeys,preprintnumbers,amsmath,amssymb]{revtex4}
 \usepackage{dcolumn}
 \usepackage{bm}

\begin{document}

 \title{ Holographic description of three dimensional
 G\"{o}del black hole }

 \author{ Ran Li }

 \thanks{Electronic mail: liran.gm.1983@gmail.com}

 \affiliation{Department of Physics,
 Henan Normal University, Xinxiang 453007, China}

 \begin{abstract}

 Three dimensional G\"{o}del black hole is
 a solution to Einstein-Maxwell-Chern
 -Simons theory with a negative cosmological
 constant. We have studied the hidden conformal symmetry
 for massive scalar field without any additional condition in the
 background of three dimensional
 non-extremal and extremal G\"{o}del black holes.
 This conformal symmetry is uncovered by the observation
 that the radial wave equations
 in both cases can all be rewritten in the form of
 $SL(2, R)$ Casimir operators through introducing
 two sets of conformal coordinates to write the $SL(2, R)$ generators.
 At last, we give the holographic dual descriptions of
 Bekenstein-Hawking entropies of
 non-extremal and extremal black holes from
 Cardy formula of conformal field theory.

 \end{abstract}



 \maketitle
 \newpage

 \section{Introduction}

 In recent years, the proposal of Kerr/CFT dual
 reveals the intriguing connection between the
 rotating black holes and two dimensional
 conformal field theory. It was initiated by Guica,
 Hartman, Song and Strominger \cite{GTSS}, who
 analysized the asymptotic symmetry property of
 near horizon geometry of extremal Kerr black hole
 (NHEK) \cite{BH} by using the approach
 of Brown and Henneaux \cite{Brown86}.
 Imposing the appropriate boundary condition
 at spatial infinity of NHEK geometry,
 the conserved charges associated with the
 asymptotic symmetry group were found to
 constitute a copy of Virasoro algebra with
 central charge proportional to angular momentum
 of black hole. So, it was conjectured that
 extremal Kerr black hole is holographically
 dual to two dimensional chiral conformal field theory.
 For more works on generalizations and other
 realizations of the proposal of extremal
 Kerr/CFT dual, one can refer to\cite{extremal,otherrealization}.

 Recently, Castro, Maloney and Strominger
 \cite{Castro} found a hidden $SL_L(2, R)\times SL_R(2, R)$
 conformal symmetry for non-extremal Kerr
 black hole through studying massless scalar field
 propagating in the near-region, which gives
 the evidence that non-extremal Kerr black hole
 may also be described by two dimensional
 conformal field theory. The essential observation
 is that the radial wave equation for scalar field
 in near-region can be reproduced by $SL_L(2, R)\times SL_R(2, R)$
 Casimir operator. However, this hidden
 $SL_L(2, R)\times SL_R(2, R)$ symmetry is only locally
 defined and is spontaneously broken to $U_L(1)\times U_R(1)$
 symmetry due to the periodic identification of
 angular coordinates, from which one can read off
 the left and the right temperatures of
 conjectured dual conformal field theory.
 Then, the dual assumption is supported
 by exactly matching the macroscopic Bekenstein-Hawking
 entropy and the microscopic entropy computed
 by Cardy formula. More recently, Chen, Long and Zhang \cite{chenlongzhang}
 have studied the hidden conformal symmetry of
 extremal black hole.
 By introducing a new set of conformal coordinates
 to write the $SL(2, R)$ generators, they are able to
 reproduce the Laplacian of scalar field in many
 extremal black holes from the $SL(2, R)$ quadratic Casimir.
 Some related works on the hidden conformal symmetry
 are listed in \cite{nonextremal,extremalhcs}.
 One can also refer to \cite{reviewkerrcft} for a comprehensive
 review on Kerr/CFT dual.

 In fact, it was noticed earlier in \cite{cveticprd} that
 the wave equation for a massless scalar field
 probing a general black hole background can be
 sufficiently simplified when certain terms are
 removed, meanwhile an $SL(2, R)^2$ symmetry emerges.
 Recently, Cvetic and Larsen \cite{cvetriclarsen} have found
 the geometrical counterpart to the omission of terms
 violating conformal symmetry in the wave equation and
 constructed the subtracted geometry corresponding to
 the wave equation exhibiting conformal symmetry.
 However, the geometrical interpretation of this
 symmetry remains obscure. In other words,
 the precise meaning of hidden conformal symmetry
 is not fully understood. So it may be worth proving
 whether the hidden conformal symmetry is captured
 by the more general black hole spacetimes.

 In this paper, we will investigate the hidden conformal symmetry
 for three dimensional non-extremal and extremal G\"{o}del
 black holes alone the lines of \cite{Castro} and \cite{chenlongzhang}
 respectively. Three dimensional G\"{o}del spacetime \cite{banados}
 is an exact solution to Einstein-Maxwell theory
 with a negative cosmological constant and a Chern-Simons term.
 This theory can be viewed as a lower dimensional toy model for
 the bosonic part of five dimensional supergravity theory.
 Three dimensional G\"{o}del black holes display the same
 peculiar properties as their higher dimensional
 counterparts\cite{banados}. The rotating black hole
 solutions on the G\"{o}del background in the context
 of five-dimensional supergravity theory have attracted
 a lot of attention \cite{godelsolution}.
 More recently, the quasinormal modes and stability
 of five-dimensional rotating G\"{o}del black holes are
 investigated by Konoplya et.al. \cite{godelqnm}.

 This paper is arranged as follows. In Sec. II,
 we give a brief review of three dimensional G\"{o}del
 black hole and its asymptotic symmetry algebra previously
 investigated by Compere and Detournay \cite{compere}.
 In Sec. III, we consider the hidden conformal symmetry
 of massive scalar field in the non-extremal black hole case.
 Firstly, it is shown that the radial wave equation can be
 explicitly solved by hypergeometric function. Secondly,
 after introducing a set of conformal coordinates,
 the radial wave equation can be rewritten in the form of $SL(2, R)$ Casimir.
 At last, the black hole entropy can be reproduced by
 combining the central charge and the left and the right temperature
 of dual conformal field theory. In Sec. IV,
 the hidden conformal symmetry of the extremal case is considered.
 The conclusion and discussion are appeared in Sec. V.

 \section{Three Dimensional G\"{o}del black hole}

 In this section, we will give a brief review of
 geometric and thermodynamic properties of
 three dimensional G\"{o}del black hole. The action of
 Einstein-Maxwell-Chern-Simons theory in 2+1 dimensions
 with a negative cosmological constant is given by
 \begin{eqnarray}
 I&=&\frac{1}{16\pi G}\int d^3 x\left[
 \sqrt{-g}\left( R+\frac{2}{l^2}
 -\frac{1}{4}F_{\mu\nu}F^{\mu\nu}\right)
 \right.\nonumber\\&&~~~~~~~~~~~~~~~~~~~~~~~~~~~~\left.
 -\frac{\alpha}{2}\epsilon^{\mu\nu\rho}
 A_{\mu}F_{\nu\rho}\right]\;.
 \end{eqnarray}

 The three dimensional G\"{o}del black hole \cite{banados}
 is an exact solution to the equations of motion derived
 from the action. The metric and gauge potential are given by
 \begin{eqnarray}
 ds^2&=&(dt-2\alpha r d\varphi)^2
 -\Delta(r) d\varphi^2 +\frac{dr^2}{\Delta(r)}\;,\\
 A_{\varphi}&=&-\frac{4GQ}{\alpha}
 +\sqrt{1-\alpha^2 l^2}\frac{2r}{l}\;,
 \end{eqnarray}
 with the metric function
 \begin{eqnarray}
  \Delta(r)=(1+\alpha^2 l^2)\frac{2r^2}{l^2}
 -8G\nu r+\frac{4GJ}{\alpha}\;.
 \end{eqnarray}
 The two parameters $\nu$ and $J$ are two integral constants in the metric function,
 which may be related to mass and angular momentum of black hole.
 Note that because of the presence of a nontrivial gauge field,
 the asymptotic geometry of three dimensional G\"{o}del
 black hole does not behave as neither de Sitter nor anti-de Sitter.

 The black hole has two horizons, i.e. the inner and the outer
 event horizons $r_{\pm}$, which are determined by the equation
 \begin{eqnarray}
 (1+\alpha^2 l^2)\frac{2r^2}{l^2}
 -8G\nu r+\frac{4GJ}{\alpha}=0\;.
 \end{eqnarray}
 The solutions give the locations of event horizons
 \begin{eqnarray}
 r_{\pm}=\frac{l^2}{1+\alpha^2 l^2}
 \left[ 2G\nu\pm\sqrt{4G^2\nu^2-
 \frac{2GJ}{\alpha}\frac{(1+\alpha^2 l^2)}{l^2}}
  \right]\;.
 \end{eqnarray}
 The outer and the inner event horizons are the coordinate
 singularities of the metric, which can be eliminated by
 a proper coordinates transformation.

 Now, Let us discuss the thermodynamics of the black hole.
 The Hawking temperature $T_H$, the angular momentum of
 the event horizon $\Omega_H$ and the Bekenstein-Hawking entropy
 $S_{BH}$ can be computed by using the standard procedure, which
 are given as
 \begin{eqnarray}
 T_H&=&\frac{(1+\alpha^2 l^2)}{4\pi\alpha l^2}
 \frac{(r_+-r_-)}{r_+}\;,\nonumber\\
 \Omega_H&=&\frac{1}{2\alpha r_+}\;,\\
 S_{BH}&=&\frac{\pi\alpha r_+}{G}\;.\nonumber
 \end{eqnarray}

 The asymptotic symmetry algebra of this spacetime
 has been studied in \cite{compere}, which turns out to be
 the semi-direct sum of the diffeomorphisms on the circle
 with two loop algebras. The covariant Poisson bracket of
 the conserved charges associated with the generators of
 asymptotic symmetry group is shown to be centrally extended to
 the semi-direct sum of a Virasoro algebra and two $u(1)$ affine algebras.
 The central charge of the Virasoro algebra is given by
 \begin{eqnarray}
  c=\frac{3\alpha l^2}{(1+\alpha^2 l^2)G}\;.
 \end{eqnarray}

 However, the black hole entropy have not been completely
 explained from the conformal field theory side \cite{compere}.
 In the following two sections, we will try to
 give a conformal field theory description of
 three dimensional G\"{o}del black hole and reproduce
 the Bekenstein-Hawking entropies of the extremal and
 non-extremal G\"{o}del black holes by combining
 the central charge (8) and the left and the right temperatures
 obtained via studying the hidden conformal symmetry of
 the probed massive scalar field.

 \section{hidden conformal symmetry: the non-extremal case}

 In this section, we will study the hidden conformal symmetry
 of the massive scalar field in the background of three dimensional
 non-extremal G\"{o}del black hole. We consider the equation of motion
 for scalar field perturbation, which is given by the Klein-Gordon equation
 \begin{eqnarray}
 \frac{1}{\sqrt{-g}}\partial_\mu\left(
 \sqrt{-g}g^{\mu\nu}\partial_{\nu}\Phi
 \right)-\mu^2\Phi=0\;.
 \end{eqnarray}
 By expanding the scalar field
 $\Phi(t, r, \varphi)=e^{-i\omega t+im\varphi}R(r)$,
 one can get the radial wave equation after some algebra
 \begin{eqnarray}
 \Delta\frac{d}{dr}\left(\Delta\frac{d}{dr} R(r)\right)
 +\left(\omega^2(4\alpha^2 r^2-\Delta)\right.&&\nonumber\\
 \left.-4\omega m \alpha r +m^2
 -\mu^2\Delta \right)R(r)=0\;&,&
 \end{eqnarray}
 where, for latter convenience, we can rewrite
 the function $\Delta(r)=\lambda(r-r_+)(r-r_-)$ with
 $\lambda=2(1+\alpha^2 l^2)/l^2$.

 Firstly, we try to show that the radial equation
 can be analytically solved by the hypergeometric function.
 For this aim, it is convenient to introduce the variable $z$ as
 \begin{eqnarray}
 z=\frac{r-r_+}{r-r_-}\;.
 \end{eqnarray}
 Then, the radial wave equation can be rewritten
 in the form of hypergeometric equation
 \begin{eqnarray}
 z(1-z)\frac{d^2 R(z)}{dz^2}+
 (1-z)\frac{dR(z)}{dz}\nonumber\\
  +\left( \frac{A}{z}+B+\frac{C}{1-z} \right)R(z)=0\;,
 \end{eqnarray}
 where the parameters $A$, $B$ and $C$ are given by
 \begin{eqnarray}
 A&=&\frac{\left(2\alpha r_+\omega-m\right)^2}{\lambda^2(r_+-r_-)^2}
 \;,\nonumber\\
 B&=&-\frac{\left(2\alpha r_-\omega-m\right)^2}{\lambda^2(r_+-r_-)^2}
 \;,\\
 C&=&\frac{4\alpha^2\omega^2}{\lambda^2}-\frac{\omega^2+\mu^2}{\lambda}\;.\nonumber
 \end{eqnarray}

 Then, the solution of radial wave equation with
 the ingoing boundary condition is given explicitly
 by the hypergeometric function
 \begin{eqnarray}
 R(z)=z^{\alpha_s}(1-z)^{\beta_s}F(a_s,b_s,c_s,z)\;,
 \end{eqnarray}
 where
 \begin{eqnarray}
 \alpha_s=-i\sqrt{A}\;,\;\;\;
 \beta_s=\frac{1}{2}-\sqrt{\frac{1}{4}-C}\;,
 \end{eqnarray}
 and
 \begin{eqnarray}
 c_s&=&2\alpha_s+1\;,\nonumber\\
 a_s&=&\alpha_s+\beta_s+i\sqrt{-B}\;,\\
 b_s&=&\alpha_s+\beta_s-i\sqrt{-B}\;.\nonumber
 \end{eqnarray}

 So, we have shown that the equation of motion
 for the massive scalar field perturbation in the background of
 three dimensional G\"{o}del black hole can be exactly solved
 in terms of hypergeometric function
 after the partial wave decomposition.
 As hypergeometric functions transform in representations of
 $SL(2, R)$, this implies the existence of a hidden conformal symmetry.
 Now we will show that the radial equation can also be
 obtained by using of the $SL(2, R)$ Casimir operator.

 Generally, in order to investigate the hidden conformal symmetry
 of wave equation, the near-region limit should be considered.
 In most of the previous papers, it is generally believed that
 the near-region is where the conformal structure appears.
 Two particular cases were reported in Ref.\cite{warpedads}
 and \cite{rlepjc}, where the low frequency limit
 for the warped $AdS_3$ black hole and the small angular limit
 for the self-dual warped $AdS_3$ black hole were found
 to probe the hidden conformal symmetry.
 However, it is pointed out in \cite{chenlongzhang} that the low frequency limit
 and the small angular limit are redundant.
 Moreover, for the warped $AdS_3$ black holes,
 the hidden conformal symmetry exist in the whole spacetime,
 which gives support to the warped AdS/CFT correspondence.
 For the three dimensional G\"{o}del black hole, we find that
 the hidden conformal symmetry can be probed by the scalar field
 without any additional condition.

 The radial equation (10) can be transformed as
 \begin{eqnarray}
  &&\left[\partial_r\left((r-r_+)(r-r_-)\partial_r\right)
  +\frac{(2\omega\alpha r_+-m)^2}{\lambda^2(r-r_+)(r_+-r_-)}\right.\nonumber\\
  &&\;\;\;\;\left.-\frac{(2\omega\alpha r_--m)^2}{\lambda^2(r-r_-)(r_+-r_-)}\right]R(r)
  \nonumber\\
  &&\;\;\;\;=\left(\frac{\mu^2}{\lambda}+\frac{\omega^2}{\lambda}
  -\frac{4\alpha^2\omega^2}{\lambda^2}\right)R(r)\;.
 \end{eqnarray}
 We will observe that this radial wave equation
 can be rewritten in the form of $SL(2, R)$ Casimir.
 It should be noted that the right-hand side is closely related to the
 conformal weights of scalar field.

 We find the appropriate conformal coordinates are given by
  \begin{eqnarray}
 w^+&=&\sqrt{\frac{r-r_+}{r-r_-}}e^{2\pi T_R\varphi}\;,\nonumber\\
 w^-&=&\sqrt{\frac{r-r_+}{r-r_-}}e^{2\pi
 T_L\varphi+2n_L t}\;,\\
 y&=&\sqrt{\frac{r_+-r_-}{r-r_-}}e^{\pi(T_L+T_R)\varphi+n_L
 t}\;,\nonumber
 \end{eqnarray}
 with the parameters
 \begin{eqnarray}
 T_R&=&\frac{\lambda}{4\pi}(r_+-r_-)\;,\nonumber\\
 T_L&=&\frac{\lambda}{4\pi}(r_++r_-)\;,\\
 n_L&=&-\frac{\lambda}{4\alpha}\;.\nonumber
 \end{eqnarray}
 Then we can locally define the vector fields
 \begin{eqnarray}
 H_1&=&i\partial_+\;,\nonumber\\
 H_0&=&i(w^+\partial_++\frac{1}{2}y\partial_y)\;,\\
 H_{-1}&=&i(w^{+2}\partial_++w^+y\partial_y-y^2\partial_-)\;,\nonumber
 \end{eqnarray}
 and
 \begin{eqnarray}
 \bar{H}_1&=&i\partial_-\;,\nonumber\\
 \bar{H}_0&=&i(w^-\partial_-+\frac{1}{2}y\partial_y)\;,\\
 \bar{H}_{-1}&=&i(w^{-2}\partial_-+w^-y\partial_y-y^2\partial_+)\;.\nonumber
 \end{eqnarray}
 These vector fields obey the $SL(2, R)$ Lie algebra
 \begin{eqnarray}
 [H_0,H_{\pm 1}]=\mp iH_{\pm 1}\;,\;\;[H_{-1},H_1]=-2iH_0\;,
 \end{eqnarray}
 and similarly for $(\bar{H}_0,\bar{H}_{\pm1})$.
 The $SL(2, R)$ quadratic Casimir operator is
 \begin{eqnarray}
 \mathcal{H}^2=\bar{\mathcal{H}}^2&=&-H_0^2+\frac{1}{2}(H_1H_{-1}+H_{-1}H_1)\nonumber\\
 &=&\frac{1}{4}(y^2\partial_y^2-y\partial_y)+y^2\partial_+\partial_-\;.
 \end{eqnarray}
 In terms of the $(t,r,\varphi)$ coordinates, the
 $SL(2, R)$ generators are given by
 \begin{eqnarray}
 H_1&=&ie^{-2\pi T_R\varphi}\left[
 \sqrt{(r-r_+)(r-r_-)}\partial_r
 \right.\nonumber\\
 &&\left.+\frac{1}{4\pi T_R}
 \frac{((r-r_+)+(r-r_-))}{\sqrt{(r-r_+)(r-r_-)}}\partial_\varphi
 \right.\nonumber\\
 &&\left.+\frac{2\alpha}{\lambda}\frac{T_L}{T_R}
 \frac{(r(r_++r_-)-2r_+r_-)}{(r++r_-)\sqrt{(r-r_+)(r-r_-)}}\partial_t
 \right]\;,\nonumber\\
 H_0&=&i\left[\frac{1}{2\pi T_R}\partial_\varphi
 +\frac{2\alpha}{\lambda}\frac{T_L}{T_R}\partial_t\right]\;,\\
 H_{-1}&=&ie^{2\pi T_R\varphi}\left[
 -\sqrt{(r-r_+)(r-r_-)}\partial_r
 \right.\nonumber\\
 &&\left.+\frac{1}{4\pi T_R}
 \frac{((r-r_+)+(r-r_-))}{\sqrt{(r-r_+)(r-r_-)}}\partial_\varphi
 \right.\nonumber\\
 &&\left.+\frac{2\alpha}{\lambda}\frac{T_L}{T_R}
 \frac{(r(r_++r_-)-2r_+r_-)}{(r++r_-)\sqrt{(r-r_+)(r-r_-)}}\partial_t
 \right]\;,\nonumber
 \end{eqnarray}
 and
 \begin{eqnarray}
 \bar{H}_1&=&ie^{-(2\pi T_L\varphi-\frac{\lambda}{2\alpha}t)}\left[
 \sqrt{(r-r_+)(r-r_-)}\partial_r
 \right.\nonumber\\
 &&\left.-\frac{1}{4\pi T_R}
 \frac{(r_+-r_-)}{\sqrt{(r-r_+)(r-r_-)}}\partial_\varphi
 \right.\nonumber\\
 &&\left.-\frac{2\alpha}{\lambda}\frac{r}{\sqrt{(r-r_+)(r-r_-)}}\partial_t
 \right]\;,\nonumber\\
 \bar{H}_0&=&-ir_0\partial_t\;,\\
 \bar{H}_{-1}&=&ie^{2\pi T_L\varphi-\frac{\lambda}{2\alpha}t}\left[
 -\sqrt{(r-r_+)(r-r_-)}\partial_r
 \right.\nonumber\\
 &&\left.-\frac{1}{4\pi T_R}
 \frac{(r_+-r_-)}{\sqrt{(r-r_+)(r-r_-)}}\partial_\varphi
 \right.\nonumber\\
 &&\left.-\frac{2\alpha}{\lambda}\frac{r}{\sqrt{(r-r_+)(r-r_-)}}\partial_t
 \right]\;,\nonumber
 \end{eqnarray}
 and the $SL(2, R)$ quadratic Casimir operator becomes
 \begin{eqnarray}
 \mathcal{H}^2&=&\partial_r((r-r_+)(r-r_-))\partial_r-
 \frac{(2\alpha r_+\partial_t+\partial_\varphi)^2}{\lambda^2(r-r_+)(r_+-r_-)}
 \nonumber\\&&
 +\frac{(2\alpha r_-\partial_t+\partial_\varphi)^2}{\lambda^2(r-r_-)(r_+-r_-)}\;.
  \end{eqnarray}
 So, for the scalar field without any additional condition,
 the wave equation can be rewritten as
 \begin{eqnarray}
 \mathcal{H}^2\Phi=\bar{\mathcal{H}}^2\Phi=
 \left(\frac{\mu^2}{\lambda}+\frac{\omega^2}{\lambda}
  -\frac{4\alpha^2\omega^2}{\lambda^2}\right)\Phi\;,
 \end{eqnarray}
 which gives the conformal weights of scalar field as
 \begin{eqnarray}
  h_L=h_R=\sqrt{\frac{1}{4}+\frac{\mu^2}{\lambda}+\frac{\omega^2}{\lambda}
  -\frac{4\alpha^2\omega^2}{\lambda^2}}-\frac{1}{2}\;.
 \end{eqnarray}

 Until now, we have uncovered the hidden $SL_L (2, R)\times SL_R (2, R)$
 symmetry of the three dimensional non-extremal G\"{o}del black hole.
 Moreover, this symmetry exists in the whole G\"{o}del spacetime,
 rather than just the Near-region in the Kerr black hole case.
 So, it is reasonable to conjecture that
 the three dimensional non-extremal G\"{o}del black hole is holographically dual
 to a conformal field theory.

 It is worth noting that hidden conformal symmetry
 is the symmetry of solution space for scalar field wave equation
 but not the one of spacetime geometry.
 However, by studying the scalar field wave equation,
 we can learn about the underlying conformal field theory
 conjectured to provide a holographic description of three
 dimensional G\"{o}del black hole.

 Recently, for the five dimensional
 asymptotically flat black hole, the subtracted geometry
 where the conformal symmetry emerges has been found
 in \cite{cvetriclarsen}. As we have shown for the three dimensional
 G\"{o}del black hole, the subtracted geometry
 is just the spacetime geometry itself, which
 has an asymptotic $SL(2, R)$ symmetry.
 This clues to a connection between the
 asymptotic $SL(2, R)$ symmetry of spacetime geometry
 and the hidden conformal structure of scalar field.
 However, the precise meaning of subtracted geometry
 is not fully understood. One can refer to \cite{cveticrecent}
 for recent progresses in this aspect.

 As a check of the conjecture,
 we want to calculate the microscopic entropy of the dual
 conformal field theory, and compare it with the
 Bekenstein-Hawking entropy of the non-extremal G\"{o}del black hole.
 Firstly, it should be noted that this hidden conformal symmetry is only
 locally defined and is spontaneously broken to $U_L (1)\times U_R (1)$
 symmetry because of the periodic identification in the $\varphi$ coordinate.
 The broken of the conformal symmetry leads to the left
 temperature $T_L$ and the right temperature $T_R$ of dual
 conformal field. Secondly, we can observe that this
 conformal symmetry is a Virasoro algebra without central
 charge. We conjecture that the central charge (8) of
 Virasoro algebra in the asymptotic symmetry will keep
 valid when studying the hidden conformal symmetry.
 So, the microscopic entropy of the dual conformal field
 theory can be computed by the Cardy formula
 \begin{eqnarray}
  S_{CFT}=\frac{\pi^2}{3}c(T_L+T_R)=\frac{\pi\alpha r_+}{G}=S_{BH}\;,
 \end{eqnarray}
 which shows the precise matching of the macroscopic
 Bekenstein-Hawking entropy and the microscopic conformal field theory entropy.

 \section{hidden conformal symmetry: the extremal case}

 In this section, we will study the hidden conformal symmetry
 of extremal G\"{o}del black hole. For the extremal case,
 the radial wave equation (10) of massive scalar field
 without any additional condition
 can be rewritten in the form of
  \begin{eqnarray}
  &&\left[\partial_r(r-r_+)^2\partial_r
  +\frac{(2\alpha r_+\omega-m)^2}{\lambda^2(r-r_+)^2}\right.\nonumber\\
  &&\;\;\;\;\left.+\frac{4\alpha\omega(2\alpha r_+\omega-m)}{\lambda^2(r-r_+)}
  \right]R(r)\nonumber\\
  &&\;\;\;\;=\left(\frac{\mu^2}{\lambda}+\frac{\omega^2}{\lambda}
  -\frac{4\alpha^2\omega^2}{\lambda^2}\right)R(r)\;.
 \end{eqnarray}

 When studying the hidden conformal symmetry
 of extremal black hole case, the conformal coordinates
 transformation (18) does not make sense because the coordinate $y$ is
 simply zero and not well-defined. Following \cite{chenlongzhang},
 we introduce the conformal coordinates
 \begin{eqnarray}
 w^+&=&\frac{1}{2}\left( \beta_1\varphi
 -\frac{\gamma_1}{r-r_+} \right)\;,\nonumber\\
 w^-&=&\frac{1}{2}\left( e^{2\pi T_L\varphi
 -\frac{\lambda}{2\alpha}t}
 -\frac{2}{\gamma_1}\right)\;,\\
 y&=&\sqrt{\frac{\gamma_1}{2(r-r_+)}}e^{\pi T_L\varphi
 -\frac{\lambda}{4\alpha}t}\;,\nonumber
 \end{eqnarray}
 with
 \begin{eqnarray}
  T_L=\frac{\lambda}{2\pi}r_+\;,\;\;\;\frac{\beta_1}{\gamma_1}=\lambda\;.
 \end{eqnarray}
 Then, the previously defined $SL(2, R)$ generators in Eq.(20) and (21) are given by
 \begin{eqnarray}
  H_1&=&i\frac{4\alpha}{\lambda\beta_1}\left(
  2\pi T_L\partial_t+\frac{\lambda}{2\alpha}\partial_\varphi
  \right)\;,\nonumber\\
  H_0&=&i\left[-(r-r_+)\partial_r
  +\frac{2\alpha\varphi}{\lambda}\left(
  2\pi T_L\partial_t+\frac{\lambda}{2\alpha}\partial_\varphi
  \right) \right]\;,\nonumber\\
  H_{-1}&=&i\left[-\beta_1\varphi(r-r_+)\partial_r
  +\frac{2\alpha\gamma_1}{\lambda(r-r_+)}\partial_t
  \right.\\
  &&\left.+\frac{\alpha}{\lambda\beta_1}
  \left(\beta_1^2\varphi^2+\frac{\gamma_1^2}{(r-r_+)^2}\right)
  \right.\nonumber\\ &&\left.
  \times\left( 2\pi T_L\partial_t+\frac{\lambda}{2\alpha}\partial_\varphi
  \right) \right]\;,\nonumber
 \end{eqnarray}
  and
 \begin{eqnarray}
 \bar{H}_1&=&2ie^{-2\pi T_L\varphi+\frac{\lambda}{2\alpha}t}
 \left[(r-r_+)\partial_r
 -\frac{2\alpha}{\lambda}\partial_t\right.\nonumber\\
 &&\left.
 -\frac{2\alpha}{\lambda^2(r-r_+)}\left(
 2\pi T_L\partial_t+\frac{\lambda}{2\alpha}\partial_\varphi
 \right)\right]\;,\nonumber\\
 \bar{H}_0&=&i\left[-\frac{2}{\gamma_1}e^{-2\pi T_L\varphi+\frac{\lambda}{2\alpha}t}
 (r-r_+)\partial_r\right.\nonumber\\
 &&\left.
 -\frac{2\alpha}{\lambda}\left(1-
 \frac{2}{\gamma_1}e^{-2\pi T_L\varphi+\frac{\lambda}{2\alpha}t}\right)
 \partial_t\right.\\
 &&\left.
 +\frac{4\alpha}{\lambda\beta_1}e^{-2\pi T_L\varphi+\frac{\lambda}{2\alpha}t}
 \left( 2\pi T_L\partial_t+\frac{\lambda}{2\alpha}\partial_\varphi
 \right)\right]\;,\nonumber\\
 \bar{H}_{-1}&=&i\left[-\frac{1}{2}\left(
 e^{2\pi T_L\varphi-\frac{\lambda}{2\alpha}t}
 -\frac{4}{\gamma_1^2}e^{-2\pi T_L\varphi+\frac{\lambda}{2\alpha}t}\right)
 \right.\nonumber\\ &&\times
 (r-r_+)\partial_r
 -\frac{\alpha}{\lambda\beta_1}\left(e^{2\pi T_L\varphi-\frac{\lambda}{2\alpha}t}
 -\frac{4}{\gamma_1}\right.\nonumber\\
 &&\left.+\frac{4}{\gamma_1^2}
 e^{-2\pi T_L\varphi+\frac{\lambda}{2\alpha}t}\right)
 \partial_t-\frac{\alpha}{\lambda^2}\left(
 e^{2\pi T_L\varphi
 -\frac{\lambda}{2\alpha}t}\right.\nonumber\\
 &&\left.\left.
 +\frac{4}{\gamma_1^2}e^{-2\pi T_L\varphi+\frac{\lambda}{2\alpha}t}\right)
 \left( 2\pi T_L\partial_t+\frac{\lambda}{2\alpha}\partial_\varphi
 \right) \right]\;.\nonumber
 \end{eqnarray}
 The Casimir operator is given by
 \begin{eqnarray}
 \mathcal{H}^2&=&\partial_r\left((r-r_+)^2\partial_r\right)
 -\frac{(2\alpha r_+\partial_t+\partial_\varphi)^2}
 {\lambda^2(r-r_+)^2}\nonumber\\
 &&
 -\frac{4\alpha(2\alpha r_+\partial_t
 +\partial_\varphi)\partial_t}{\lambda^2(r-r_+)}\;.
 \end{eqnarray}

 Actually, there exist one degree of freedom to define the
 conformal coordinates (31) without affect the form of the
 Casimir operator, i.e. such degree of freedom does not
 change the underlying physics.

 So, once again, the radial wave equation for the scalar field
 without any additional condition
 in the background of extremal G\"{o}del
 black hole can be rewritten as
 \begin{eqnarray}
 \mathcal{H}^2\Phi=\bar{\mathcal{H}}^2\Phi
  =\left(\frac{\mu^2}{\lambda}+\frac{\omega^2}{\lambda}
  -\frac{4\alpha^2\omega^2}{\lambda^2}\right)\Phi\;.
 \end{eqnarray}

 This result indicates that there exist the hidden conformal
 symmetry for three dimensional extremal G\"{o}del black hole.
 Similar to the hidden conformal symmetry of non-extremal black holes,
 the vector fields are not globally defined. The periodic
 identification along the $\varphi$ coordinate breaks
 this symmetry. If we conjecture that the extremal G\"{o}del
 black hole is dual to two dimensional conformal field theory,
 the breaking of hidden conformal symmetry leads to the
 non-vanishing left temperature $T_L$ and the vanishing right
 temperature $T_R$ of the dual conformal field theory.
 In other words, for the case of extremal black hole,
 only the left sector in the dual
 conformal field theory is excited. This result agrees with the
 one of the non-extremal case in the extremal limit
 (see Eq.(19) in Sec.III).

 Now, we are in a position to check the dual conjecture
 by matching the macroscopic entropy from the gravity side
 and the microscopic entropy from the conformal theory side.
 For the extremal case, the microscopic entropy comes entirely from the
 left sector and takes the form
 \begin{eqnarray}
 S_{CFT}=\frac{\pi^2}{3}c T_L=\frac{\pi\alpha r_+}{G}=S_{BH}\;,
 \end{eqnarray}
 which gives the microscopic explanation of Bekenstein-Hawking
 entropy.

 \section{Conclusion and Discussion}

 In this paper, we have studied the hidden conformal symmetry
 of massive scalar field for the three dimensional non-extremal
 and extremal black holes. The conformal symmetries are
 uncovered by the observations that the radial wave equations
 for the non-extremal case and the extremal case can be
 rewritten in the form of $SL(2, R)$ Casimir through introducing
 two sets of conformal coordinates transformations
 to write the $SL(2, R)$ generators, respectively.
 This conformal symmetry implies a dual connection
 between G\"{o}del black hole and conformal field theory.
 At last, the dual conjectures are checked by
 reproducing the black hole Bekenstein-Hawking entropy from
 the Cardy formula via combining the central charge and
 the left and the right temperature
 of dual conformal field theory.

 Generally, the dual conjecture implied by the hidden conformal
 symmetry can also be checked by matching the two-point correlation
 function or the absorption probability from the gravity side and
 conformal field theory side.
 However, the examination is hard to perform for
 the present case due to the difficulties in defining the
 conserved quantities for the three-dimensional G\"{o}del
 black hole. It can be seen from the metric function (4),
 the black hole has three parameters $\nu$, $J$ and $Q$.
 It has been shown in \cite{banados} that,
 via the rigorous definition of conserved charges
 and tensor calculation, the parameter $\nu$ is
 the conserved quantity associated to the
 killing vector $\partial_t$.
 However, it is observed in \cite{compere} that,
 under the change of coordinates $r\rightarrow-r$,
 $\phi\rightarrow-\phi$, the solutions with
 the parameters $(\nu, J, Q)$ can be changed to
 the solutions with the parameters $(-\nu, J, -Q)$.
 So the conserved quantity $\nu$ does not provide
 a satisfactory definition of black hole mass.
 The first law of thermodynamics for the three
 dimensional G\"{o}del black hole is still missing in the
 literatures. So, in the case of lacking the satisfactory
 definition of conserved quantities and the first law
 of thermodynamics, it is unable to deduce the conjugate
 charges associated with the left and the right
 temperatures in the conformal field theory side,
 i.e. one can not obtain the absorption probability
 from the conformal field theory side, which make the
 compare from the two sides no sense.
 This aspect can be explored in the future.

 \section*{Acknowledgements}

 I would like to thank Pu-Jian Mao and Ming-Fan Li for reading the manuscript and useful
 comments. This work was supported by NSFC, China (Grant No. 11147145).

 \end{document}